\documentclass{emulateapj}
\slugcomment{Draft Version, Accepted for publication in ApJ}
\usepackage{times}
\usepackage{bm}
\usepackage{color}

\newcommand{\bolB}{{\bm  B}}

\newcommand{\bolV}{{\bm  V}}

\newcommand{\beq}{\begin{equation}}
\newcommand{\eeq}{\end{equation}}
\newcommand{\beqn}{\begin{eqnarray}}
\newcommand{\eeqn}{\end{eqnarray}}
\newcommand{\beqno}{\begin{equation*}}
\newcommand{\eeqno}{\end{equation*}}
\newcommand{\beqnno}{\begin{eqnarray*}}
\newcommand{\eeqnno}{\end{eqnarray*}}
\newcommand{\etal}{{\em et al. }}

\shorttitle{A Magnetohydrodynamic Model of the M87 Jet. II.}
\shortauthors{Nakamura \& Meier}

\begin{document}

\title{A Magnetohydrodynamic Model of The M87 Jet. II. Self-consistent
Quad-shock Jet Model \\ for Optical Relativistic Motions and Particle
Acceleration}

\author{Masanori Nakamura\altaffilmark{1} \& David L. Meier\altaffilmark{2}}
\affil{
$^1$Institute of Astronomy \& Astrophysics, Academia Sinica, 11F of
Astronomy-Mathematics Building, AS/NTU No. 1, Taipei 10617, Taiwan; 
nakamura@asiaa.sinica.edu.tw}
\affil{
$^2$Jet Propulsion Laboratory, California Institute of Technology, 
Pasadena, CA 91109, USA; david.l.meier@jpl.nasa.gov}

\begin{abstract}
We describe a new paradigm for understanding both relativistic motions
and particle acceleration in the M87 jet: a magnetically dominated
relativistic flow that naturally produces four relativistic
magnetohydrodynamic (MHD) shocks (forward/reverse fast and slow
modes). We apply this model to a set of optical super- and subluminal
motions discovered by Biretta and coworkers with the {\em Hubble Space
Telescope} during 1994 -- 1998. The model concept consists of ejection
of a {\em single} relativistic Poynting jet, which possesses a coherent
helical (poloidal + toroidal) magnetic component, at the remarkably
flaring point HST-1.  We are able to reproduce quantitatively proper
motions of components seen in the {\em optical} observations of HST-1
with the same model we used previously to describe similar features in
radio VLBI observations in 2005 -- 2006. This indicates that the quad
relativistic MHD shock model can be applied generally to recurring pairs
of super/subluminal knots ejected from the upstream edge of the HST-1
complex as observed from radio to optical wavelengths, with
forward/reverse fast-mode MHD shocks then responsible for observed
moving features. Moreover, we identify such intrinsic properties as the
shock compression ratio, degree of magnetization, and magnetic obliquity
and show that they are suitable to mediate diffusive shock acceleration
of relativistic particles via the first-order Fermi process.  We suggest
that relativistic MHD shocks in Poynting-flux dominated helical jets may
play a role in explaining observed emission and proper motions in many
AGNs.
\end{abstract}

\keywords{galaxies:individual: M87 --- galaxies: active --- galaxies:
jets --- methods: numerical --- MHD}

\section{Introduction}
\label{sec:int}

In this paper we apply our previous relativistic MHD shock model for the
2005 M87 radio jet \citep[hereafter Paper I]{N10} to the {\em optical}
super/subluminal knots discovered by \citet[]{B99} using the {\em Hubble
Space Telescope} ({\em HST}) at five epochs between 1994 and 1998. These
observations reveal superluminal features in the range $5c$ -- $6c$ with
some subluminal components located around 0$\farcs$8--1.6$\arcsec$
(projected) from the core (or $\sim260$ -- 520 pc de-projected for a
viewing angle of $\sim 14^{\circ}$; Wang \& Zhou 2009). This region has
been named as the ``HST-1'' complex. So far HST-1 is one of the most
energetic elements of the M87 jet, exhibiting both fast and slow
(super/subluminal) motions as well as the birth of new components and
the fading of older ones \citep[]{B99, C07}. The global structure of the
jet is characterized as a parabolic stream on the sub-arcsecond scale,
which changes into a conical stream beyond one arcsecond; HST-1 is
indeed the narrow ``neck'' in the jet, indicating an over-collimated
focal point (or ``recollimation shock'') \citep[]{AN12, NA13}.

Multi-band light curves of HST-1 reveal an impulsive flare event that
had a peak in 2005 \citep[]{H06, C07, M09}. As reported in
\citet[]{C07}, between 2005 December and 2006 February, the component
HST-1c, which had been ejected during 2004 -- 2005 from HST-1d (the
upstream edge in the HST-1 complex), {\em split into two bright
features}: a faster moving component (c1: $4.3c \pm 0.7c$) and a slower
moving one (c2: $0.47c \pm 0.39c$). The ejection of these components is
believed to be associated with the HST-1 flare occurring in 2005. The
simultaneous rise and fall of light curves at all wavelengths (radio,
optical, NUV, and X-ray bands) indicate that the flare was a local event
caused by a simple compression at HST-1 \citep[]{H06, H09}, which
created an increase of the synchrotron particle energy at all
wavelengths equally and a fractional polarization in the optical band at
a level from 20\% to 40\% \citep[]{P11}.

Furthermore, the very high energy (VHE) $\gamma$-ray emission in the TeV
band that occurred in 2005 \citep[]{A06} may be associated with
contemporaneous radio-to-X-ray flaring of HST-1, while the nucleus
itself was in a {\em quiescent} phase from radio to X-ray bands during
the $\gamma$-ray flare event \citep[]{A12}. The VLBA monitoring at 22/43
GHz of EGRET blazars has established a statistical association that
$\gamma$-ray flares at high levels occur shortly after ejections of new
superluminal components of parsec-scale jets in nearby VLBI cores
\citep[]{J01}. Thus, we suggest that the VHE flare associated with the
superluminal knot ejection in M87 is intrinsically similar to events
seen in other blazars.

Paper I proposed a model to explain the ejected super/subluminal VLBA
knots from HST-1d in 2005 -- 2006 \citep[]{C07} as a pair of
forward/reverse fast-mode MHD shocks in a strongly magnetized
relativistic flow that possesses an ordered helical field component.  A
simple test of this model would be to find another appropriate candidate
quad shock complex in the M87 jet. Here, we seek it in the earlier HST
observations of 1994 -- 1998 \citep[]{B99}, and we suggest that
HST-1$\epsilon$ ($6.00c \pm 0.48c$) / HST-1 East ($0.84c \pm 0.11c$) in
their observations are a similar pair to HST-1c2/c1. With several
moderate changes in model parameters, we then reproduce the component
motions with our quad MHD shock model and show that the shock conditions
there are ideal for particle acceleration. This paper is organized as
follows. In \S 2, we outline the numerical model. In \S 3, we describe
our numerical results. Discussions and conclusions are given in \S 4.

\section{Numerical and Physical Model}
\label{sec:Model}

\subsection{Component Geometry and Emission}

A detailed description of our model concept ({\em a magnetically
dominated relativistic flow}) is in \S 2 of Paper I. Using a linear
scale of 78 pc arcsec$^{-1}$ \citep[$D=16$ Mpc;][]{T91}, a proper motion
of 1 mas yr$^{-1}$ at M87 corresponds to an apparent velocity of
0.25$c$. In \citet[]{B99}, it is suggested that the most upstream
component of the HST-1 complex (HST-1 East, at 870 mas from the core and
moving relatively slowly at $0.84c \pm 0.11c$) had given birth to at
least three superluminal components. In 1995, HST-1 East appeared to
eject a new component at 870 mas from the core (HST-1$\epsilon$: $6.00c
\pm 0.48c$).

{\em However}, we suggest a different scenario based on the later
observations of \citet[]{C07}. In their VLBA observations, HST-1d is the
dominant feature in the HST-1 complex in the early epochs before
2005. Between 2005 and 2006, the location of HST-1d is basically {\em
stationary} to within $\sim$ 2 mas (i.e., its motion is $<0.25 c$) at
860 mas from the core. Then, the radio knot HST-1c must have {\em
emerged from} HST-1d in the downstream direction ($>$ 860 mas). Since
the upstream region of the HST-1 complex seems to be well resolved in
VLBA (but not so well resolved in HST) observations, we thus consider
HST-1 East to be a moving component ejected from the stationary HST-1d
(the upstream edge in the HST-1 complex).

Following Paper I, we assert that the lateral gas compression at HST-1
(and its expansion after maximum squeezing) causes the ejection of new
shock components. Very recently, \citet[]{L13} suggested the de Laval
nozzle-like shape as an explanation for the multi-wavelength light
curves during the 2005 flaring event at HST-1 described above; an
adiabatic compression/expansion of the flow cross section may be
responsible for observed multi-wavelength synchrotron light curves
\citep[e.g.,][]{H06, H09}. Also, as we have described in Paper I,
axisymmetric non-relativistic and relativistic MHD numerical simulations
of strongly magnetized, super-fast magnetosonic flows with a helically
twisted magnetic field component produce a magnetic chamber, which opens
and closes intermittently, ejecting multiple quad shock components into
the downstream "Nose Cone" region \citep[]{L89, K99}.

During the past few decades, an extensive monitoring of the M87 jet
downstream from HST-1 ($1 - 18 \arcsec$ or $0.1 - 1.5$ kpc in
projection) has been conducted in a wide range of wavelengths at radio
(VLA), optical (HST), and X-ray (Chandra) bands. Emissions from radio to
X-ray bands resemble each other in morphology, indicating a common
synchrotron radiation process \citep[]{M02, WY02, PW05}. The
observations also suggest that {\it in situ} particle acceleration at
shocks \citep[via the first-order Fermi process;][]{BO78} occurs in the
large scale M87 jet, as evidenced both from the electron lifetime scale
(much shorter than the jet travel time from the nucleus) and from
spectral fits to the broadband spectra.  A relativistic particle energy
distribution $n(E) dE \propto E^{-\delta} dE$ would need a spectral
index steeper than $\delta=2$ in order to produce the radio through
optical to X-ray synchrotron spectrum in the M87 jet (at HST-1 and its
downstream region); synchrotron models have been fit with $\delta=2.2$
at all energies and all locations along the jet \citep[]{PW05} and with
about $\delta=2.36$ on average \citep[]{LS07}. These agree very well
with the conditions needed for diffusive shock acceleration (DSA)
[$\delta=2-2.5$; e.g., \citet[]{KD01, RBRD07, SBDB12}].

Intensity profiles, which are taken across the jet (FWHM) at knot A for
HST and VLA observations, yield motions of 1$\arcsec$.13 in the radio
band and only 0$\arcsec$.85 in the optical \citep[]{S96}. Note that
average intensity profiles normal to the jet axis for knots D, E, F, I,
B, and C also have a similar tendency; optical knots are more compact
and centrally concentrated than the radio knots. Comparison between the
optical and radio polarimetry by \citet[]{P99} provides additional
evidence that optical- and radio-emitting electrons are not completely
co-located. Their results show that the degree of polarization varies
less in the radio than in the optical, indicating that the
optical-emitting electrons are located closer to the jet axis, whereas
most of the radio-emitting electrons are located nearer the jet surface.

\subsection{Numerical Setup}

The basic numerical treatment is essentially the same as in Paper I
(see, \S 3 and 4).  We solve the special relativistic MHD (SRMHD)
equations in a cylindrical 1.5-dimensional approximation (axisymmetry in
the azimuthal direction $\phi$) along the $z$-axis at a fixed
cylindrical radius $r$.  Our normalization details are summarized in
Table \ref{tbl:unit}.  Compared to Paper I for modeling HST-1c2/c1
\citep[]{C07}, here we consider several moderate changes regarding the
initial conditions for modeling HST-1$\epsilon$/East \citep[]{B99}.  By
assuming a viewing angle $\theta_{\rm v} \sim 14^{\circ}$ at HST-1
\citep[]{WZ09}, a maximum ``intrinsic'' speed (including an error) in
HST observations of the M87 jet can be estimated from the ``apparent''
speed of the fastest moving component HST-1$\epsilon$: $\beta_{\rm app}
= 6.00 \pm 0.48$ \citep[]{B99} with $\beta_{\rm int}=\beta_{\rm
app}/(\beta_{\rm app} \cos \theta_{\rm v}+\sin \theta_{\rm v})\simeq
0.992$, where $\beta \equiv V/c$.  It is assumed that the intrinsic
speed is associated with the jet fluid speed $\beta_{\rm fl}=V_{\rm
fl}/c$, with $\beta_{\rm int} \lesssim \beta_{\rm fl}$
\citep[]{B95}\footnote{Readers can refer to the related argument in
section 2 of \citet[]{NA13} concerning the possibility that the observed
proper motions are correlated with the underlying bulk flow in AGN
jets.}.

We model the jet as a highly magnetized medium with low plasma-$\beta$
values (a ratio of the gas pressure to the magnetic pressure)
$\beta_{\rm p} < 0.1$ and small magnetic obliquity angles $\sim
10^{\circ}$ (measured in the rest frame of the fluid). The jet is
injected as a trans-fast magnetosonic, relativistic flow (Lorentz
factor: $\gamma \simeq 11.48$) into a medium flowing with a
sub-relativistic speed ($\gamma \simeq 1.07$).  Under these conditions,
the jet naturally produces a set of four relativistic MHD shocks in the
system. The computational domain $z \in [-0.04,\,2.0]$ (parsec in a
dimensional scale), which is resolved with 5100 grid points, assigns two
uniform states (up: upstream and down: downstream) separated at
$z=0.0$. Time integration is followed until $t=2.0$ ($\sim$ 6.4 yr) to
examine an early phase of relativistic MHD shock propagation and inspect
the individual wave fronts.

The following initial conditions are prescribed:
\begin{eqnarray}
&&(\rho,\,V_{\phi},\,V_{z},\,B_{\phi},\,B_{z},\,p)^{\rm up}=
 (1.0,\,0,\,0.996,\,8.0,\,3.0,\,0.405) \nonumber
\end{eqnarray}
on the upstream side $(-0.04 \leq z \leq 0.0)$ and
\begin{eqnarray}
&&(\rho,\,V_{\phi},\,V_{z},\,B_{\phi},\,B_{z},\,p)^{\rm down}=
 (1.0,\,0,\,0.360,\,0.7,\,3.0,\,0.052) \nonumber
\end{eqnarray}
on the downstream side. (The numerical time integration uses a CFL
number of 0.5.) The main difference between initial conditions in the
present paper and in Paper I is the implementation of {\em two} uniform
states of $B_{\phi}$ on each side (measured in the rest frame of the
galaxy), while Paper I specifies one uniform state on both sides. This
treatment will affect mainly the compression ratio $r_{\rm cmp}$ at the
forward fast-mode shock (as is discussed in \S 5 of Paper I). Note that
$B_{\phi}^{\rm up}/B_{\phi}^{\rm down} \simeq 1.07$ (quasi-uniform state
on both sides) in current initial conditions, if we measure in the rest
frame of the fluid.

\begin{figure}[ht]
\begin{center}
\includegraphics[scale=0.45, angle=0]{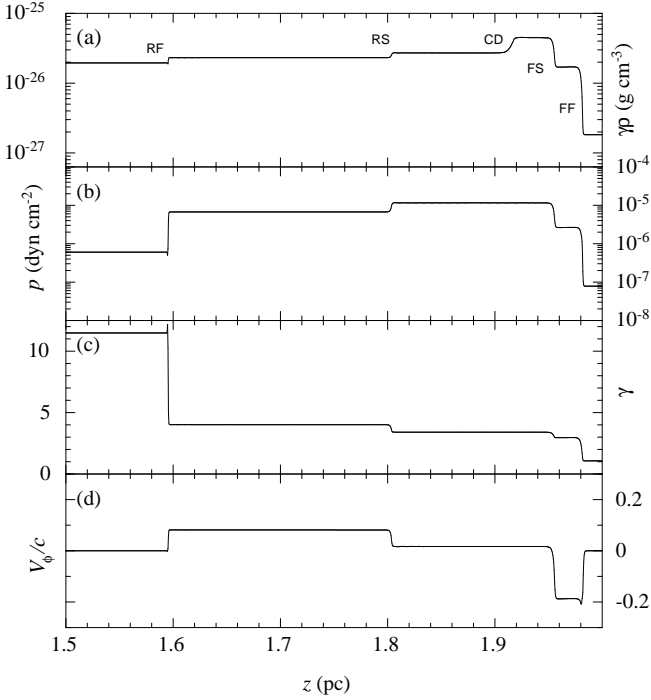}
\caption{
\label{fig: f1} 
(a) -- (d) $\log (\gamma \rho)$, $\log (p)$, $\gamma$, and $V_{\phi}/c$,
respectively, shown at $t = 2.0$.  Only the region $1.5 \leq z \leq 2.0$
is displayed.  Note that panels (a) -- (d) are measured in the rest
frame of the galaxy. Each discontinuity is labeled in (a).}
\end{center}
\end{figure}

\begin{figure}
\begin{center}
\includegraphics[scale=0.45, angle=0]{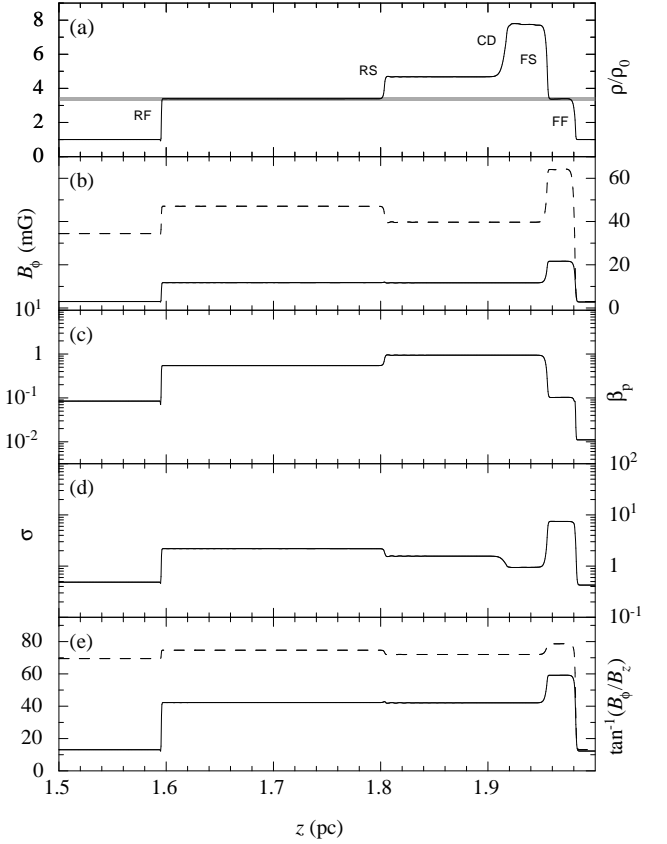}  
\caption{\label{fig: f2} 
(a) -- (e) $\rho/\rho_{0}$, $B_{\phi}$, $\log (\beta_{\rm p})$
(plasma-$\beta$), $\log (\sigma)$ (magnetization parameter), and
$\theta$ (magnetic obliquity angle), respectively, shown at
$t = 2.0$.  Only the region $1.5 \leq z \leq 2.0$ is displayed.  Note
that all quantities in each panel are measured in the rest frame of the
fluid (drawn as a solid line), with the exception of the broken
lines in (b) and (e), which are measured in the rest frame of the
galaxy.  Each discontinuity is labeled in (a).  A gray shaded area in
(a) indicates a proper compression ratio ($r_{\rm cmp}\sim 3.3$--3.5)
for a relativistic DSA.}
\end{center}
\end{figure}

\section{Numerical and Physical Results}
\label{sec:Results}

\subsection{Jet Flow and Shock Propagation}

Figure \ref{fig: f1} shows the propagation of the relativistic MHD wave
fronts; snapshots of various quantities at $t=2.0$ are illustrated in
the rest frame of the galaxy. The distribution of proper density $\gamma
\rho$ shows the quad MHD shock pattern plus a contact discontinuity (CD
or entropy wave), all with constant speeds. While all features move
downstream in the galaxy frame, {\em in a reference frame that
co-moves with the jet plasma near the CD}, these waves propagate in both
the forward (F) and reverse (R) directions. Here we adopt the convention
of counting shocks beginning with the one farthest from the origin of
the disturbances (HST-1). Two of the four shocks, the first and the
fourth, are forward fast-mode (FF) and reverse fast-mode (RF) shocks,
respectively. The other two, the second and the third, are forward
slow-mode (FS) and reverse slow-mode (RS) shocks.

Basic features shown in Fig. \ref{fig: f1} are similar to those in
Fig. 3 of Paper I. The flowing gas is compressed twice across the first
(FF) and second (FS) shocks, while it is expanded in crossing the third
(RS) and last (RF) shocks respectively, as seen in (a) and (b).  As a
result, the gas pressure at the accumulated region between FS and RS
shocks (RS-CD-FS) increases by almost two orders of magnitude compared
to the pre-shocked region by twice compressions at the FF and FS
shocks. As one moves from large to small $z$, $\gamma$ increases with
gradual steps in the first, second, and third shocks, and greatly
increases in the last shock to the injection level $\gamma \simeq 11.48$
shown in (c). From (d), $V_{\phi}$ changes as well at each shock
discontinuity; the region FF-FS and the region RS-RF are {\em
counter-rotating} when viewed from a frame that rotates with the plasma
near the CD, as was also seen in Paper I.

Strengths and propagation speeds of the four shocks remain constant with
distance as they propagate in our coordinate system: axial propagation
($z$-direction) in a uniform medium (constant sound and Alfv\'en speeds)
in a fixed-radius cylindrical shell. Individual speeds of shock fronts
are estimated as $V_{\rm FF} \sim 0.99c$, $V_{\rm FS} \sim 0.98c$,
$V_{\rm RS} \sim 0.90c$, and $V_{\rm RF} \sim 0.80c$, respectively. For
a viewing angle of $\theta \sim 14^{\circ}$ at HST-1 \citep[]{WZ09}, the
faster component HST-1$\epsilon$ has $\sim 0.99c$, while the slower
component HST-1 East has $\sim 0.79c$. As is mentioned in Paper I, a
separation of observed super/subluminal components can be identified as
distinct proper motions of two fast-mode MHD shocks (FF/RF), instead of
two slow-mode MHD shocks due to an ineffectiveness of the DSA in
slow-mode shocks \citep[e.g.,][]{KD99}. Thus, our numerical model is
consistent with observations \citep[]{B99}.

\subsection{Particle Acceleration}

In order to examine the efficiency of high energy particle acceleration
by the DSA, {\em i.e.}  the first-order Fermi process \citep[]{BO78}, we
show several quantities in Fig. \ref{fig: f2} that are measured in the
rest frame of the fluid. Panel (a) shows the shock compression ratio
$r_{\rm cmp}$ of the density $\rho$ to the ambient value
$\rho_{0}$. This ratio at each shock front is $\sim 3.4$ at both the FF
and RF, $\sim 2.3$ at the FS, and $\sim 1.4$ at the RS.

For the DSA process in non-relativistic shocks (shock propagation speed
$V_{s} \ll c$), the spectral slope $\delta$ in a power-law distribution
of the relativistic particle energy does not depend on the details of
the flow (the magnetic field orientation near the shock, the mechanism
of particle diffusion, or other microscopic physics involved). Instead,
$\delta$ depends only on the compression ratio $r_{\rm cmp}$
\citep[e.g.,][]{B78} as
\beqn
\delta\equiv\frac{r_{\rm cmp}+2}{r_{\rm cmp}-1},
\eeqn
where $\delta \simeq 2$ corresponds 
to strong shocks with a maximum compression ($r_{\rm cmp} \simeq 4$)
\citep[]{D83, BE87}.

However, in the case of relativistic shocks ($V_{s} \sim c$), $\delta
\sim 2.2$--2.3 is expected \citep[][]{W97, BO98, K00, A01},
corresponding to $r_{\rm cmp} \sim 3.3$--3.5. So, our numerical result
of $r_{\rm cmp} \sim 3.4$ at the FF/RF is entirely consistent with the
expected value for relativistic DSA theory and observations
\citep[]{PW05, LS07}. Note that an efficiency of the particle
acceleration by the relativistic DSA crucially depends on background
conditions, such as both the magnetization and magnetic obliquity of the
upstream plasma.  Furthermore, $\delta \sim 2.2$--2.3 may be valid only
in quasi-parallel (small magnetic obliquity) shocks, while a large
departure from this range is confirmed in Monte Carlo simulations
\citep[e.g.,][]{BO98, NO06}.

Panel (b) of Fig. \ref{fig: f2} shows the distribution of $B_{\phi}$ in
the rest frame of the fluid as well as the galaxy. It increases
across the first shock, decreases across the second one, increases again
across the third shock, and finally decreases across the fourth one.  In
the rest frame of the galaxy, the azimuthal field component is much
larger than the axial field component $\sim 10$ mG, while 
in the rest frame of the fluid both are comparable. A field of $\sim 10$ mG 
near the HST-1 complex has been derived from variability time scales in
optical and X-ray observations \citep[]{P03}.  From panel (c) of
Fig. \ref{fig: f2}, we find the gas pressure near the CD is in approximate
equipartition with the magnetic pressure in the rest frame of the fluid
($\beta_{\rm p} \sim 1$, as was also seen in Paper I).

Using the definition in \citet[]{N11}, the magnetization parameter
$\sigma$ in the local plasma rest frame is defined as the ratio of the
Poynting flux to the matter energy flux:
\beqn
\sigma \equiv \frac{B_{\phi}^2}{4\pi\gamma^2 \rho c^2}.
\eeqn
We also define the obliquity angle $\theta$ in the
local plasma rest frame:
\beqn
\theta \equiv \tan^{-1} \left(\frac{B_{\phi}}{\gamma B_{z}}\right).
\eeqn

Recent 2.5D/3D particle-in-cell simulations \citep[e.g.,][]{S08, SS09} confirm
that particle acceleration is mostly mediated by the DSA process for
quasi-parallel field ($\theta \lesssim 10^{\circ}$), but shock drift
acceleration (SDA) is the main acceleration mechanism for larger, yet
still {\em subluminal} \citep[in {\em de Hoffmann-Teller}
frame:][]{dHT50} magnetic obliquity. The critical angle for the shock to
be ``subluminal''\footnote{If the shock is ``superluminal'', it is
difficult to for the DSA process to proceed \citep[][]{BK90}.} decreases
with increasing upstream bulk Lorentz factor $\gamma$ and magnetization
$\sigma$ but stays confined within a relatively narrow range
($\theta_{\rm crit} \simeq$ 26$^{\circ}$-- 42$^{\circ}$) for moderate
magnetization ($\sigma \lesssim 1.0$) \citep[]{SS09}. Panel (d) and (e)
of Fig. \ref{fig: f2} show the distribution of $\sigma$ in the rest
frame of the fluid and $\theta $ in both the fluid rest and
galaxy frames. We can see $\sigma \lesssim 0.5$ and $\theta \lesssim
13^{\circ}$ upstream of both the FF and RF, indicating the DSA process may be
feasible in both quasi-parallel shocks FF/RF.

In our model, the quad shock system is initiated at HST-1 and propagates
in a conical streamline. As is shown in Paper I, $B_{\phi}$ becomes much
more dominant than $B_{z}$ in the downstream direction. Furthermore,
study of proper motions indicate a systematic deceleration of
propagating knots \citep[]{B95, B99, M13}. By combining these aspects,
$\theta$ eventually becomes large, indicating a quasi-perpendicular
shock. In order to maintain a universal value $\delta \sim 2.2$--2.3 for
the relativistic DSA, large amplitude MHD turbulence
($\kappa_{\perp}/\kappa_{\parallel} \simeq 1$, where $\kappa_{\perp}$
and $\kappa_{\parallel}$ are the cross-field and the parallel diffusion
coefficient, respectively) near the shock would be required \citep[{\em
otherwise}, $\delta$ can be much steeper than the above asymptotic
values in the absence of large turbulence, e.g., ][]{BO98, OB02, NO04,
N06}. Note that $\kappa_{\perp}$ and $\kappa_{\parallel}$ are in units
of $c r_{\rm g}$, where $r_{\rm g}$ is the particle gyration radius in
the unperturbed background field. Therefore, it may be beyond our scope,
but recent relativistic MHD simulations of mildly relativistic shocks
with $V_{\rm s}\sim 0.4 c$--$0.9 c$ suggest that perpendicular shocks
produce highly turbulent field amplification in the postshock region
\citep[]{M11}.

Finally we remark on the efficiency of shock dissipation in highly
magnetized (Poynting-flux dominated) relativistic flows. \citet[]{K12}
found that the dissipation efficiency (ratio of thermal to total energy
flux densities) of a fast magnetosonic shock is still a quite high
fraction, $\sim$ 30\% ($\sigma=1.0$) -- 80\% ($\sigma=0.1$) of the total
energy flux. This is the case mainly because only the kinetic energy is
dissipated, and it represents only a small fraction of the total energy
flowing through the shock. We therefore propose that our quad
relativistic shock model may explain not only the relativistic bulk
motions in a pair of super/subluminal features in AGN jets, but also
the particle acceleration that takes place in them.

\section{Discussion and Conclusions}
\label{sec:Dis-Con}

The basic assumption of our model posits ejection of a {\em single}
relativistic jet, which naturally produces four MHD shocks, from a
stationary feature (standing over-collimation Mach disk / oblique shock
system) in compact radio sources that produce a pair of super/subluminal
knots.  In M87, we believe that the HST-1 complex is the place where
these events occur \citep[]{B99,C07}.  Very recently, \citet[]{G12}
reported two superluminal components ejected from HST-1 after 2007
(component 2 in 2008 and component 3 in 2010). In their analysis,
component 2 is identified as being similar to HST-1c (seen in
\citet[]{C07}); it eventually splits into two sub-components, although
the authors argue that the slow sub-component may be an underlying,
standing or very slowly moving feature (a detailed proper motion
analysis was not conducted for this sub-component). {\em However}, we
suggest that component 2 may represent the ejection of a third quad
relativistic shock system in the M87 jet, which possesses both sub-
(reverse) and superluminal (forward) features.  \citet[]{G12} also
pointed out simultaneous timings between the superluminal component
ejections and VHE flares in 2008 and 2010 \citep[]{A12}, suggesting that
structural changes at the upstream edge of HST-1 are related to these
flares.

Very recently, \citet[]{M13} studied proper motions of the M87 jet on
arcsecond (kiloparsec) scales by using more than a decade of {\em HST}
archival imaging. Significant new apparent motions $\gtrsim c$ have been
found at the knot A/B/C complex. Furthermore, knots C and A move in
opposite directions {\em transverse} to the jet axis with $V \gtrsim
0.1c$ in projection. This may indicate a counter-rotational motion
around the jet axis as expected for a pair of fast-mode shocks (FF/RF)
in an older (and now much larger) quad MHD shock system. (Such motions
occur in our current simulation of the much smaller HST-1 complex, as
seen in (d) of Fig. \ref{fig: f1} and also in Paper I.)  Overall
velocity profiles along the jet axis, as well as transverse to that
axis, may be explained as embedded flow trajectories within systematic
helical magnetic fields \citep[]{M13}. Velocity components that lie
upstream of knot A are observed to still have highly relativistic, and
thus one-sided (i.e., {\em negative}), transverse motions (Doppler
boosted towards us). Once the jet becomes mildly relativistic, however,
we are able to track the full (i.e., both {\em positive} and {\em
negative}) transverse motions of the helical pattern in projection.
Furthermore, there is a conspicuous ``tip-to-tail'' alignment of almost
all the velocity vectors within the knot A/B/C complex, strongly
suggesting a flattened view of a helical motion which might result in
such a ``zig-zag'' pattern. In the framework of a quad MHD shock system,
a pair of fast-mode shocks (FF/RF), corresponding to the knots C/A, may
be responsible for driving the helical distortion near the postshock
region of B via the current-driven helical kink ($m=1$) instability
\citep[]{NM04}. Thus, we propose that the region A/B/C may be a good
example (on the kiloparsec scale) of the interplay between the MHD
shocks and current-driven instability, where the magnetic field plays a
fundamental role in the M87 jet dynamics, as originally suggested in
Paper I.

It is widely believed that moving shocks in jets (``shock-in-jet''
model) are responsible for the synchrotron emission in blazars
\citep[{\em e.g.},][]{BK79, M80}. A subset of the preceding,
superluminal (forward shock) and the following, stationary/subluminal
(reverse shock) features are frequently seen in VLBI observations
\citep[]{J05, L09}. Among shock-in-jet models, the following two major
scenarios have been discussed in a non-MHD framework: (1) a collision of
the faster shock with either the preceding slowly moving shock
\citep[''internal shock'' model: {\em e.g.},][]{S01} or (2) a standing
shock complex \citep[{\em e.g.},][]{DM88, S04}. Note that both forward
and reverse sonic shocks are expected in these models. An extension of
the internal shock model with a perpendicular MHD forward/reverse shocks
has been performed by \citet[]{M07}. As mentioned in \S \ref{sec:int},
strong $\gamma$-ray flares occur after ejections of new superluminal
components from parsec-scale regions of jets in nearby VLBI cores
\citep[]{J01}. Instead of an internal shock scenario, we suggest here
that there is a {\em standing} shock at HST-1 based on the observational
aspects. Furthermore, because of the strong polarization associated with
the knots in M87, as well as the superluminal motion, we must model
these shocks using special relativistic magnetohydrodynamic simulations.

In this paper, we investigate a pair of super/subluminal motions in the
M87 jet based on the quad relativistic MHD shock model \citep[]{N10}.
The model concept consists of ejection at HST-1 of a {\em single}
relativistic Poynting jet, which possesses a coherent helical (poloidal
+ toroidal) magnetic component that naturally produces such features, as
a counterpart to the hydrodynamic Mach disk - oblique shock system.
HST-1$\epsilon$/East, which were identified in HST observations
\citep[]{B99}, are modeled quantitatively with one-dimensional
axisymmetric SRMHD simulations. We conclude that forward/reverse
fast-mode MHD shocks are a promising explanation for the observed
features, not only with regard to their intrinsic motions, but also in
the efficiency of the diffusive shock acceleration (through the
first-order Fermi process) of non-thermal particle accelerations at the
shock fronts.  Three fundamentals at the fast-mode MHD shocks derived
from the simulations (shock compression ratio, degree of magnetization,
and magnetic obliquity (magnetic pitch angle)) are suitable to mediate a
Fermi-I process.

While we do not yet fully investigate the hypothesis that ``all
relativistic jets are dominated by the toroidal magnetic field component
in the observer's frame'', $B_{\phi}/B_{z} \sim \gamma$ \citep[]{LPG05}
certainly holds in the interknot (intershock) region of the M87 jet as
we found ($B_{\phi}/\gamma/B_{z} \sim 1$ in the fluid frame).
Therefore, we suggest our model may be applicable to many
super/subluminal features of AGN jets in general.

M.N. acknowledges part of this research was carried out under supported
by the Allan C. Davis fellowship jointly awarded by the Department of
Physics and Astronomy at Johns Hopkins University and the Space
Telescope Science Institute. Part of this research also was carried out
at the Jet Propulsion Laboratory, California Institute of Technology,
under contract with the National Aeronautics and Space Administration.

\begin{deluxetable}{rlll}
\tabletypesize{\scriptsize}
\tablecaption{Units of Physical Quantities for Normalization. \label{tbl:unit}}
\tablewidth{0pt}
\tablehead{
\colhead{Physical Quantities} & \colhead{Description}   &
\colhead{Normalization Units} & \colhead{Typical Values}}
\startdata
$z$\dotfill        & Length         & $L_{0}$         & $3.1 \times 10^{18}$ cm (1 parsec)\\
${\bolV}$\dotfill  & Velocity field & $c$             & $3.0 \times 10^{10}$ cm s$^{-1}$\\
$t$\dotfill        & Time           & $L_{0}/c$       & $1.0 \times 10^{8}$ s (3.2 yr)\\
$\rho$\dotfill     & Density        & $\rho_{0}$      & $1.7 \times 10^{-27}$ g cm$^{-3}$\\
$p$\dotfill        & Pressure       & $\rho_{0} c^{2}$ & $1.5 \times 10^{-6}$ dyn cm$^{-2}$\\
${\bolB}$\dotfill  & Magnetic field & $\sqrt{4 \pi \rho_{0}c^2}$ & $4.3 \times 10^{-3}$ G
\enddata
\end{deluxetable}

\end{document}